\documentclass[conference]{IEEEtran}
\IEEEoverridecommandlockouts
\usepackage{cite}
\usepackage{amsmath,amssymb,amsfonts}
\usepackage{algorithmic}
\usepackage{graphicx}
\usepackage{textcomp}
\usepackage{xcolor}
\DeclareMathOperator*{\argmax}{arg\,max}
\def\BibTeX{{\rm B\kern-.05em{\sc i\kern-.025em b}\kern-.08em
    T\kern-.1667em\lower.7ex\hbox{E}\kern-.125emX}}
\begin{document}


\title{Multi-Dialect Arabic Speech Recognition}

\author{\IEEEauthorblockN{Abbas Raza Ali}
\IEEEauthorblockA{\textit{Faculty of Science and Technology} \\
\textit{Bournemouth University}\\
Poole BH12 5BB, United Kingdom \\
abbas.raza.ali@gmail.com}}

\maketitle

\begin{abstract}
This paper presents the design and development of multi-dialect automatic speech recognition for Arabic. Deep neural networks are becoming an effective tool to solve sequential data problems, particularly, adopting an end-to-end training of the system. Arabic speech recognition is a complex task because of the existence of multiple dialects, non-availability of large corpora and missing vocalization. Thus, the first contribution of this work is the development of a large multi-dialectal corpus with either full or at least partial vocalized transcription. Additionally, the open-source corpus has been gathered from multiple sources that bring non-standard Arabic alphabets in transcription which are normalized by defining a common character-set. The second contribution is the development of a framework to train an acoustic model achieving state-of-the-art performance. The network architecture comprises of a combination of convolutional and recurrent layers. The spectrogram features of the audio data are extracted in the frequency vs time domain and fed in the network. The output frames, produced by the recurrent model, are further trained to align the audio features with its corresponding transcription sequences. The sequence alignment is performed using a beam search decoder with a tetra-gram language model. The proposed system achieved a 14\% error rate which outperforms previous systems.
\end{abstract}


\section{Introduction}
Deep Neural Networks (DNN) has become a de-facto standard in the Computer Vision task. In recent years, they are gaining traction in speech and language processing tasks with a significant improvement in acoustic modelling~\cite{Amodei2015}. Despite human speech varies with respect to pitch, amplitude, duration and the combination of phonemes and morphemes, yet how they tie together to create words. The acoustic modeling is a contextual problem which generates sequential information. The Recurrent Neural Networks (RNN), a class of DNN, is considered as the most powerful tool for this kind of problem~\cite{Graves2013}. Instead of combining RNNs with Hidden Markov Model (HMM) and Gaussian Mixture Model (GMM), \cite{Awni2014} introduces a concept of end-to-end automatic speech recognition (ASR) using solely RNN. Thus, end-to-end training replaces the traditional acoustic modeling approach to predict different states efficiently. Also, it makes the overall task simplified and combined with a language model (LM) gives a state-of-the-art performance. 

Arabic is the $5^{th}$ most widely spoken language even though it is an under resource language. Although, there are few open-source Arabic speech corpora available which can be used to build a baseline speech recognition model. However, for the phonetically rich language like Arabic, these resources are not sufficient to build a reasonably accurate acoustic model. The key challenge in the case of the Arabic language is to obtain diacritized transcription of the audios which is not very common in open-source speech corpora. The native Arabic speakers understand the meaning of the text from the context rather than expecting diacritized text. Another challenge associated with the Arabic language is the existence of several dialects with a multitude of pronunciations~\cite{Ananthakrishnan2005}.

This paper addresses a number of the challenges faced during the design and development of multi-dialect Arabic ASR. Typically, the DNNs require a large amount of relevant data to train a model. Hence, a large multi-dialectal corpus has been developed which is one of the contributions of this work. Additionally, the open-source corpus has been gathered from multiple sources. The corpus brings variants of character-set and their transcripts are either fully-, partially- or un-diacritized. Thus, the transcription of all the sources is normalized to a common character-set, defined in the next section, whereas automatic diacritization followed by a manual review has been performed. The automatic diacritization systems are quite effective for Arabic, Urdu \cite{Ali2010}, and a few other languages. 

The rest of the paper is organized as follows. The development of speech corpus, from various sources, is discussed in Section~\ref{sec:Corpus}. Section~\ref{sec:Methodology} is devoted to elaborate on the overall approach of this work. Section~\ref{sec:Results} reports the experimental results followed by their analysis. Finally, the paper is concluded in Section~\ref{sec:Conclusion}.

\begin{table*}[!htbp]
\caption{List of speech corpora used to build acoustic model}
\begin{center}
\label{tab:corpora}
\begin{tabular}{|l|l|l|l|} \hline
\textbf{Dataset} & \textbf{Dialect} & \textbf{Duration} & \textbf{Prompt} \\ \hline 
Aldiri~\cite{AlDiri2004} & MSA & 2 hours & Read speech  \\ \hline
KACST~\cite{Mansour2004} & MSA & 2 hours & Read speech  \\ \hline
Isolated Words~\cite{Alalshekmubarak2014} & MSA & 10 hours & Read speech  \\ \hline
Arabic News Channel~\cite{Wray2015} & MSA, Egyptian, Gulf, Levantine, North African & 57 hours & Conversations, interviews, and news documentary  \\ \hline
KSU~\cite{Alsulaiman2013} & MSA & 120 hours & Read speech  \\ \hline
MGB-2~\cite{Khurana2016} & MSA, Egyptian, Gulf, Levantine, North Africa & 1,200 hours & Conversations, interviews, and news documentary  \\ \hline
This work & MSA, Egyptian, Gulf & 70 hours & Read speech  \\ \hline
\end{tabular}
\end{center}
\end{table*}

\section{Speech Corpus} \label{sec:Corpus}
In order to train an acoustic model, a large amount of speech corpus has been gathered. The corpus, covering multiple dialects, is recorded from ~450-500 Arabic speakers possessing different accents. The distribution of the gender is around 35-65 for female and male speakers respectively. Additionally, the corpus is a mixture of various speaking styles including a) read speech, b) answer speech (or interviews), c) descriptive speech (or documentary), and d) spontaneous speech (or conversation). However, irrespective of the speech style, the audio is segmented on average between 3 and 50 seconds which is suitable to train a good speech recognizer. The quality of the transcription varied significantly in different sources. The variations in the transcript are in the form of un-diacritized or semi-diacritized transcript. Also, the conversational speech contains overlapping of phrases and the usage of non-standard Arabic alphabets~\cite{Ali2017}.

The speech corpus is gathered from multiple sources which are listed in Table~\ref{tab:corpora}. The Aldiri corpus consists of 4,740 utterances spoken by 6 speakers with the distribution of 3 males and 3 females. The KACST dataset consists of 4,573 utterances in Modern Standard Arabic (MSA) dialect. The `Isolated words' data-source contains 9,992 utterances of 20 words spoken by 50 male Arabic speakers. The `Arabic News Channel' corpus is composed of conversational speech style with around 27,000 utterances. There are few challenges associated with this corpus because of its speaking style, such as overlapping speech and usage of non-standard alphabets. The non-standard alphabets that appear in the transcript are normalized to a common character-set and the overlapping phrases are omitted from the final corpus. The KSU database is recorded from more than 200 speakers belonging to different ethnic groups. The transcription of the corpus is composed of phonetically rich and balanced words, digits, sentences, and paragraphs. The audio is recorded from low and high-quality microphones and telephony mediums in different environments including office, sound-proof room, and cafeteria. The Multi-Genre Broadcast (MGB)-2 corpus is recorded from 19 different programs of an Arabic TV channel where the content covers three speaking styles including conversations, interviews, and documentaries. The MSA dialect was most dominating in the programs along with Egyptian, Gulf, Levantine and North African.

\subsection{Character-set}
\begin{figure*}[!htbp]
  \centering
  \includegraphics[scale = 0.75]{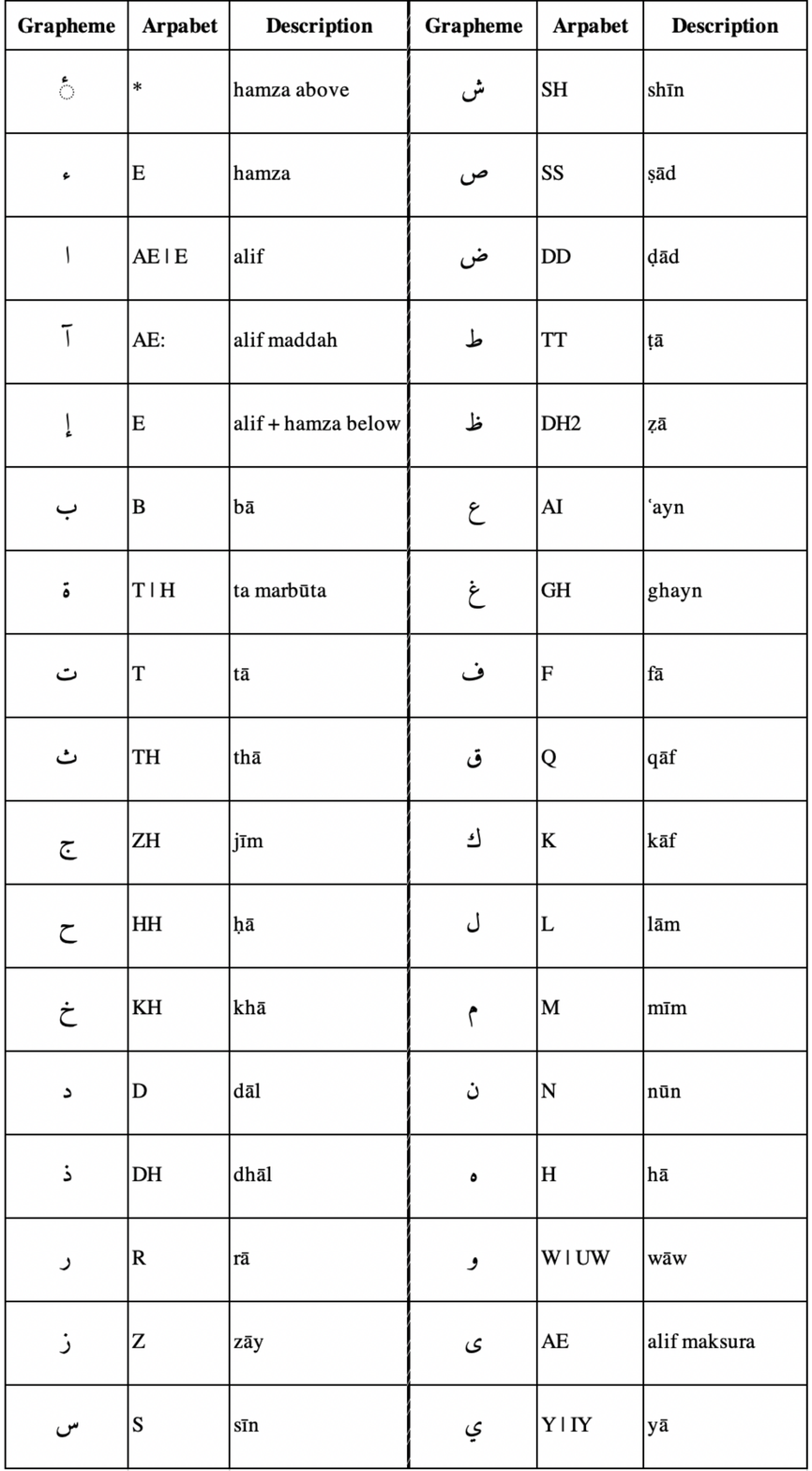} 
  \caption{Arabic character-set with phonemes and description}
  \label{fig:characterset}
\end{figure*}

The raw Arabic corpora are gathered from multiple sources using different character-sets in their transcripts. A common character-set has been defined to normalize all the sources. Initially, for some experiments,  the regular Arabic short vowels called: fatha, damma, and kasra including a diacritical-mark shadda are part of the character-set. The remaining Arabic diacritical-marks were rear in most of the data-sources, hence, they were excluded comprising of fathatan, kasratan, dammatan, and sukun. Three long vowels are part of the character-set including alef, waw and ya. The rest of the alphabets are standard consonants of Arabic. In the final experiments, the short vowels and shadda are also excluded which were causing dataset imbalance. The Arabic character-set used in training the system is shown in Fig.~\ref{fig:characterset}.  

\subsection{Data Cleansing}
The data cleansing and normalization become critical when developing a corpus from different sources that possess different rules. The audio files are recorded using the low and high quality of microphones with different background noise. A high-pass filter is applied to the audios with a cutoff frequency of 150 Hz to cancel the DC offset. The audio files are converting into a single-channel 16-bit 16,000 Hz sample rate encoded using Linear Pulse Code Modulation (\textbf{L}PCM). The long audio files are split between 2-30 seconds segments. Similarly, the transcript is normalized accordingly by segmenting long sentences into multiple phrases to synchronize them with their corresponding audios. Also, different sources bring a different combination of diacritical-marks which were normalized by, initially, limiting the text to fatha, damma, kasra, and shadda. In the later experiments, all the diacritical-marks were removed except `hamza above'. The `hamza above' mark was used with alef, waw and ya. The transcripts with non-standard Arabic characters have been completely removed from the corpus.   

\subsection{Development of New Corpus}
\begin{figure*}[!htbp]
\centering
\includegraphics[scale = 0.7]{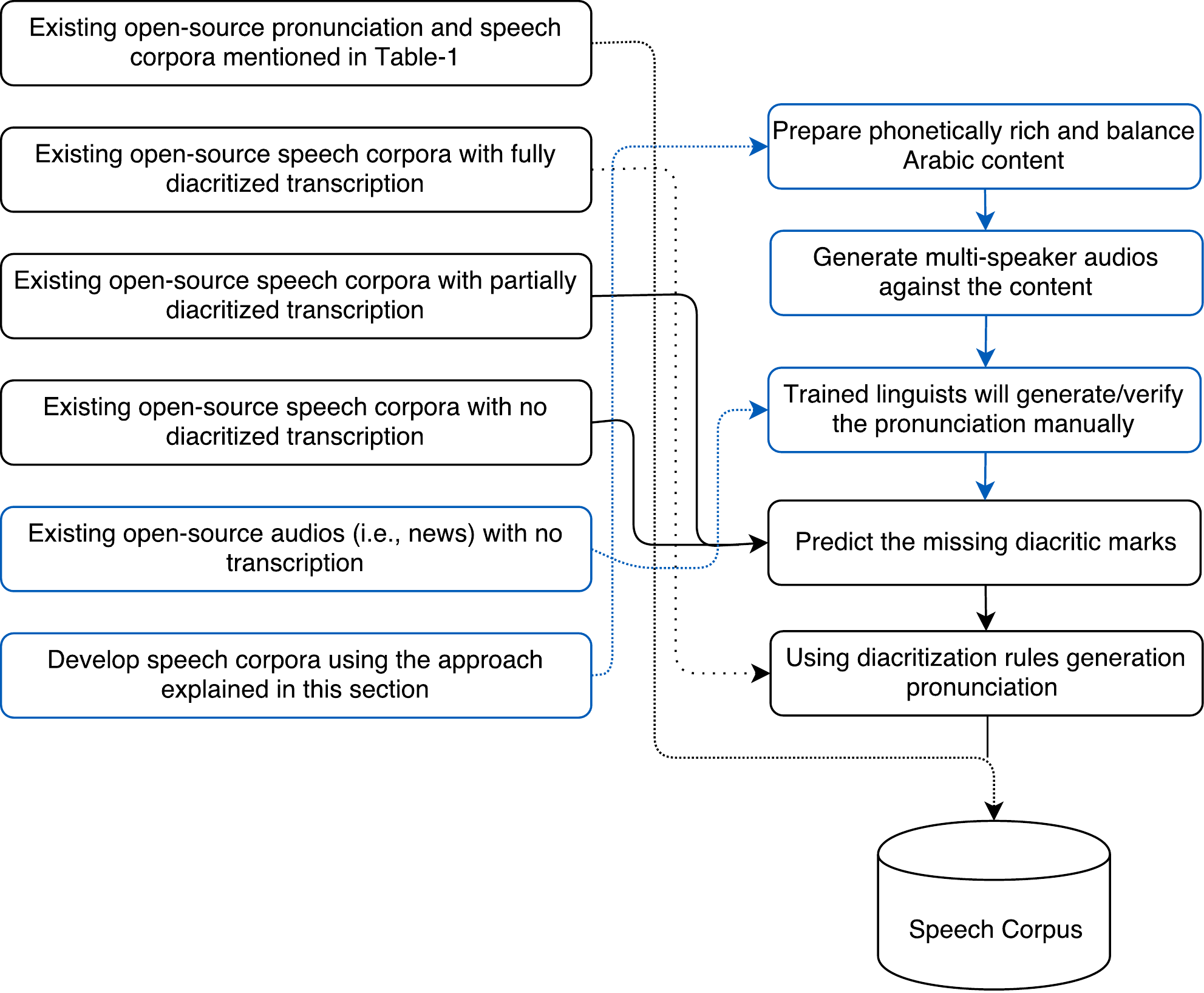}
\caption{Steps to build Arabic Speech Corpus}
\label{fig:corpus_development}
\end{figure*}

In addition to the collection of open-source Arabic speech corpora, an effort has been made to design and develop a phonetically rich and balanced corpus. The development of the corpus consists of three key stages: 1) the content preparation, 2) speech recording and validation and 3) verification. The high-level steps of collecting and developing Arabic speech corpus are shown in Fig.~\ref{fig:corpus_development}. The idea is to first utilize all the available open-source resources to build a basic corpus followed by the development of a new corpus. Following are the three key areas of new corpus development:

\subsubsection{Transcription}
A phonetically rich and balanced transcription is prepared which covers all the possibilities of Arabic phonemes and their different positions in a word. Thus, the transcript of Saudi Accented Arabic Voice Bank (SAAVB) \cite{Alghamdi2008} was chosen. The transcript includes 1,033 unique phrases including numbers, words, sentences, and paragraphs that were designed to differentiate the number of dialects.

\subsubsection{Corpus Recording and Validation}
The corpus recording process started with the recruitment of a large number of Arabic speakers possessing different dialects. The recruited speakers were required to read ~190 phrases that were assigned to them via an automatic system. A recorded phrase is first evaluated using speech processing tools and is stored in the database only if it passes the quality test. In case of failure, because of either noise, unnecessary silence, repetition, etc., the speaker is required to re-record the same phrase until it gets approved by the speech analyst. This process is repeated on all the phrases and speakers. 

\subsubsection{Verification}
A detailed verification of the recording setup has been performed by the linguists before the initiation of a recording session. The verification process is not limited to the verification of recording equipment, it also checks the quality of the recorded speech. Further, prior to the recording sessions, the speakers are instructed on the usage of recording hardware and software.

\section{Methodology} \label{sec:Methodology}
Arabic is a Semitic language having unique phonemes and phonetic features, multiple dialects, and a complicated morphological word structure~\cite{Selouani1998}. In order to train a speech recognizer using this kind of complex data, the DNN capacity has been increased via depth. Two models are trained independently namely; acoustic model and language model, see Equation~\ref{eq:1}. The speech corpus, comprising of transcribed audios, is used to train the acoustic model. The acoustic model is used to predict phonemes, given a segment of audio features. On the other hand, an n-gram based language model predicts the next word in a sequence. It trains on text corpus comprising of transcripts of conversations.  

\begin{equation} \label{eq:1}
\begin{aligned}
\textrm{ASR} &= \argmax_{w \in vocab} \textrm{ } (\textrm{language model}) \textrm{x} (\textrm{acoustic model}) \\
 &= \argmax_{w \in vocab} \textrm{ } (words) \textrm{ x } (features | words) \\
 &= \argmax_{w \in vocab} \textrm{ } (w) \textrm{ x } (x | w)
\end{aligned}
\end{equation}

\begin{figure*}[!htbp]
\centering
  \includegraphics[viewport=210 75 400 800, scale = 0.48]{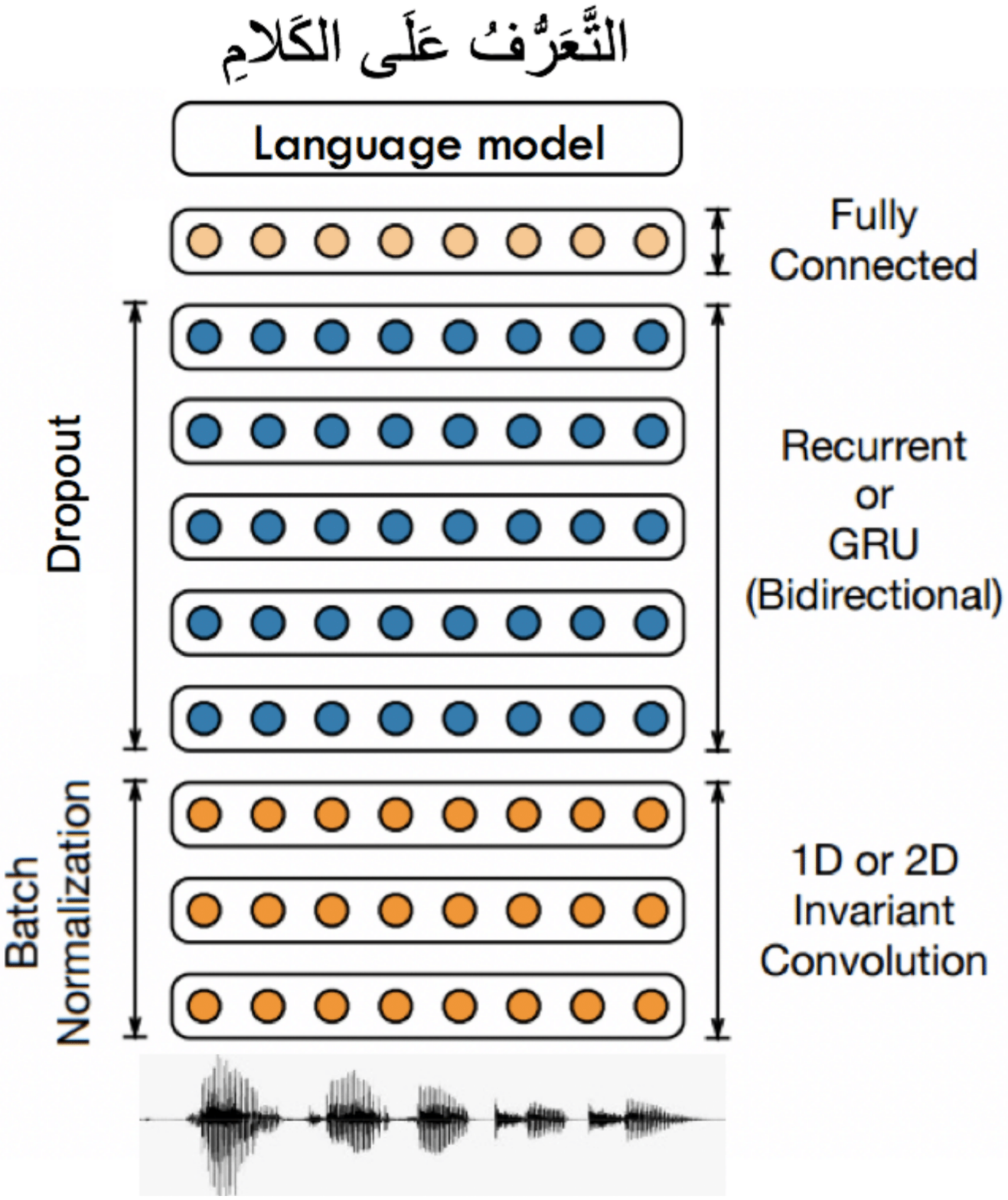}
  \caption{The architecture of end-to-end Arabic speech recognition. In the experiments 2-3 convolution and 4-5 recurrent layers have been tested.}
  \label{fig:architecture}
\end{figure*}

The network consists of a combination of convolutional and recurrent layers followed by a fully-connected layer. The convolutional layers are limited to 2-3 with a stack of 4-5 bi-directional recurrent layers. Various sizes of network width are tried including 768 and 1024. Also, two types of RNNs have experimented which are Long short-term memory (LSTM) and Gated Recurrent Units (GRUs). A clipped Rectified Linear Unit (ReLU) function has been applied as formulated in Equation~\ref{eq:2}. The End-to-End speech recognition architecture is shown in Fig.~\ref{fig:architecture}.

\begin{equation} \label{eq:2}
\sigma(x) = min(max(x, 0), 20) 
\end{equation}

Three convolutional layers have been used with 32, 32 and 96 filter sizes having dimensions 41x11, 21x11 and 21x11 respectively. The strides of 2x2, 2x1 and 2x1 have been applied to down-sample the input. The regularisation is key to better performance by avoiding over-fitting when using RNN. Hence, batch normalization (BatchNorm) is used with convolutional layers whereas a dropout of 20\% is used for recurrent layers. The BatchNorm also enhances the training speed of this kind of very deep networks with marginally improved accuracy. To avoid over-fitting, early stopping is also used while training the network. The sequence alignment is performed through the Connectionist Temporal Classification (CTC) which is found to be very successful for speech and language processing tasks~\cite{Najafian2017}. The training examples of varying lengths are sorted in increasing order of length since short examples are comparatively less costly.

The inference process applies normalization and extracts features of the new audio. The trained acoustic model produces phoneme probabilities over time. These probabilities are fed to the decoder, along with a language model, to compute the most likely sequence of words for the given audio input. 

\subsection{Feature Extraction}
The Mel-frequency Cepstral Coefficients features (MFCC) have demonstrated promising performance for speech recognition task~\cite{Bouchakour2018}. However, for this work, the spectrograms are extracted from the audio files which are fed as input to the convolutional layers of the network. The spectrogram is a two-dimensional representation of audio data consisting of frequency spectrum over time which can be processed as a two-dimensional image. The filter banks are computed over windows of 0.02 seconds with strides of 0.01 seconds~\cite{Awni2014}. 

\subsection{Acoustic Modeling}
The acoustic model is trained using more than 1,400 hours of speech data. A deep and wide network has been designed to train the acoustic model along with sequence alignment. The optimal set of hyper-parameters is found by performing a number of experiments with different values of filters and strides of convolutional layers, and the type and width of recurrent layers. The weights are initialized with Xavier-initialization~\cite{glorot2010}. A dropout of 20\% has been used for recurrent layers whereas BatchNorm is used for convolutional layers. The network is trained with Adam optimizer~\cite{Kingma2015} with a variable learning rate which decays by a factor of 10 on every alternative epoch with an initial value of 0.001. 

The recurrent networks are trained as frame-level classifiers for speech recognition tasks~\cite{Graves2014}. These frames further require training to align the input audio with corresponding output transcription sequences. The CTC allows a sequence training without requiring any prior alignment between the input and target sequences. In this work, the CTC beam search along with a trained word-level LM has been used. The LM guides the CTC beam search decoder on domain-specific and miss-spelled terms. The CTC is used with a beamwidth of 512.

\subsection{Language Model}
The word-level LM predicts the upcoming word given the previous sequence of words~\cite{Ali2018}. A word-level tetra-gram LM is trained on a large Arabic corpus. The short vowels and shadda diacritical-mark are also removed from the language model corpus to synchronize with the character-set of the acoustic model. The LM has been trained over millions of unique utterances using KenLM with Kneser-Ney smoothed tetra-gram model~\cite{Heafield2013}. The beam search is used to find the optimal word sequence. The trained model is quite effective for spelling and implicitly disambiguate homophones. 

The corpus that is used to build LM is collected from multiple sources listed in Table~\ref{table:lm_corpora}. The MGB-2 has diversity in broadcast news comprising of 19 distinct programs of an Arabic TV channel. The text is crawled from the website of the channel. The Tashkeela corpus contains 75.6 million fully vocalized terms that are crawled from 97 classical and modern Arabic books. The consolidated text is normalized using the same rules which are applied to the transcription of the audio data.

\begin{table}[!htbp]
\caption{List of corpora used to train language model}
\begin{center}
\label{table:lm_corpora}
\begin{tabular}{|l|l|l|} \hline
\textbf{Dataset} & \textbf{Vocabulary} & \textbf{Domain} \\ \hline
MGB-2 LM~\cite{Khurana2016} & 130 million & News  \\ \hline
Tashkeela~\cite{Zerrouki2017} & 75.6 million & Islamic classical books \\  \hline
\end{tabular}
\end{center}
\end{table}

\section{Analysis and Results} \label{sec:Results}
The proposed architecture is tested with different combinations of hyper-parameters using a trial-and-error approach. The corpus is divided into training, development and validation sets with a ratio of 70\%, 20\%, and 10\% respectively. The experiments are conducted in a distributed training framework for 32-35 epochs using 10 Nvidia 1080ti GPUs. The total batch size of 80 (8 batches per GPUs) is used with a learning rate of 0.0001 and a momentum of 0.99 to train multiple models. There are 8 different models trained varying along 2-3 convolutional and 4-5 recurrent layers, and RNN width (768 or 1024) and type (LSTM or GRU). The models which perform well during the training phase on the development set were chosen for further evaluation. Their performance is measured using word error rate (WER) which is formulated in Equation~\ref{eq:3}.  

\begin{equation} \label{eq:3}
\textrm{WER} = \frac{\textrm{substitutions (S)} + \textrm{deletions (D)} + \textrm{insertions (I)}}{\textrm{total words in the reference (N)}}
\end{equation}

\begin{table}[!htbp]
\begin{center}
\caption{Comparison of different arrangements of convolutional and recurrent layers along with network width and RNN type}
\label{table:accuracies}
\begin{tabular}{|l|r|l|r|} \hline
\textbf{Architecture} & \textbf{Neurons} & \textbf{RNN Type} & \textbf{WER (\%)} \\ \hline
    2-CNN, 4-RNN & 768 & GRU & \textbf{14.07}  \\  \hline
    2-CNN, 4-RNN & 768 & LSTM &  14.67  \\   \hline
    2-CNN, 4-RNN & 1024 & GRU &  15.32  \\   \hline 
    2-CNN, 4-RNN & 1024 & LSTM & 15.54 \\ \hline
    3-CNN, 5-RNN & 768 & GRU &   14.88 \\ \hline
    3-CNN, 5-RNN & 768 & LSTM &  15.30 \\ \hline
    3-CNN, 5-RNN & 1024 & GRU &  14.23 \\  \hline          
    3-CNN, 5-RNN & 1024 & LSTM & 14.61 \\  \hline
\end{tabular}
\end{center}
\end{table}

The outcome of the acoustic model is decoded using the CTC beam search with LM. The final WER of the different models is reported in Table~\ref{table:accuracies}. The model with 2 convolutional layers followed by 4 bi-directional recurrent layers, composed of GRU cell and 768 neurons, significantly outperformed the rest of the variants. The statistical significance is computed using a paired t-test of the top-performing architecture's WER with the rest of the WERs. Further, the high-performing model is using a fully-connected layer followed by a CTC beam search decoder with LM.

Table~\ref{table:comparision} provides a comparison of the proposed system with the prior research. The performance of several DNN variants, such as Feed-forward, CNN, LSTM, Time Delayed Neural Networks (TDNN) and GMM have been evaluated.

\begin{table}[!htbp]
\begin{center}
\caption{Comparison of the proposed approach with existing systems}
\label{table:comparision}
\begin{tabular}{|p{1.15cm}|l|p{2.29cm}|r|}  \hline
\textbf{Research} & \textbf{Acoustic/Lang. Model} & \textbf{Speech Corpus} & \textbf{WER \%} \\ \hline  
\cite{Peddinti15} & TDNN/3-gram & MGB-2 & 20.5 \\  \hline
\cite{AlHanai2016} & RNN/4-gram & MGB-2 & 18.3 \\   \hline
\cite{Ali2017} & RNN/3-gram & MGB2 & 14.70 \\   \hline
\cite{Menacer2017} & RNN/4-gram &  Nemlar$^{\mathrm{1}}$ \& NetDC$^{\mathrm{2}}$ & 14.42 \\ \hline
This work & CNN, RNN/4-gram & MGB-2 \& custom & \textbf{14.07} \\   \hline
\multicolumn{4}{l}{$^{\mathrm{1}}$http://catalog.elra.info/product\_info.php?products\_id=874} \\ 
\multicolumn{4}{l}{$^{\mathrm{2}}$http://catalog.elra.info/product\_info.php?products\_id=13} \\ 
\end{tabular}
\end{center}
\end{table}

\section{Conclusions} \label{sec:Conclusion}
An end-to-end automatic speech recognition system for multi-dialectal Arabic language has been presented in this paper. The key contributions include the development of large multi-dialectal corpus and an end-to-end speech recognition model with accuracy close to the state-of-the-art for the Arabic language. This accuracy is the result of numerous trials to explore various network architectures and their hyper-parameters, accompanied by several techniques including sorting of training samples, BatchNorm, early stopping and decoding with LM. 

Two key contributions are claimed in this work. The first contribution is the development of a large multi-dialectal Arabic speech corpus consisting of ~1,400 hours of speech data. The corpus has been gathered from multiple sources and normalized on a common character-set. The second contribution is the development of a framework to train the Arabic speech recognition acoustic model with close to state-of-the-art accuracy. The system has achieved an overall WER of 14\%.

\vspace{12pt}
\color{red}

\end{document}